\documentclass[a4paper,onecolumn,unpublished]{quantumarticle}
\pdfoutput=1
\PassOptionsToPackage{hyphens}{url}
\usepackage[usenames,dvipsnames]{xcolor}
\usepackage{graphicx}
\usepackage[ colorlinks = true, 
             linkcolor = violet,
             urlcolor  = ForestGreen,
             citecolor = ForestGreen,
             anchorcolor = Magenta,
]{hyperref} 
\usepackage{amsmath,amssymb,amsthm,mathtools,mathrsfs}
\usepackage{pifont}
\usepackage{authblk} 
\usepackage{bbm}
\usepackage{tikz}
\usepackage[all]{xy}
\usepackage[compat=1.1.0]{tikz-feynman}

\usepackage{dsfont}
\usepackage[all]{xy} 
\usepackage{enumerate}
\usepackage{tocbasic}

\DeclareTOCStyleEntry[
  beforeskip=0.3em plus 1pt,
  pagenumberformat=\textbf
]{tocline}{section}

\makeatletter
\renewcommand*\env@matrix[1][\arraystretch]{%
  \edef\arraystretch{#1}%
  \hskip -\arraycolsep
  \let\@ifnextchar\new@ifnextchar
  \array{*\c@MaxMatrixCols c}}
\makeatother

\theoremstyle{plain}
\newtheorem{theorem}[equation]{Theorem}
\newtheorem{lemma}[equation]{Lemma}
\newtheorem{proposition}[equation]{Proposition}
\newtheorem{corollary}[equation]{Corollary}
\theoremstyle{definition}
\newtheorem{definition}[equation]{Definition}

\newtheorem{question}[equation]{Question}

\newtheorem{problem}[equation]{Problem}
\newtheorem{example}[equation]{Example}
\newtheorem{exercise}[equation]{Exercise}
\newtheorem*{answer}{Answer}
\newtheorem*{solution}{Solution}
\newtheorem{remark}[equation]{Remark}
\newtheorem{terminology}[equation]{Terminology}

\newtheorem{notation}[equation]{Notation}
\newtheorem{noterm}[equation]{Notation and Terminology}

\newcommand\define[1]{\emph{\textbf{#1}}}

\numberwithin{equation}{section}

\newcommand{\be}{\begin{equation}}
\newcommand{\ee}{\end{equation}}
\newcommand{\bx}{\begin{example}}
\newcommand{\ex}{\end{example}}
\newcommand{\bex}{\begin{exercise}}
\newcommand{\eex}{\end{exercise}}
\newcommand{\ban}{\begin{answer}}
\newcommand{\ean}{\end{answer}}
\newcommand{\bt}{\begin{theorem}}
\newcommand{\et}{\end{theorem}}
\newcommand{\bc}{\begin{corollary}}
\newcommand{\ec}{\end{corollary}}
\newcommand{\blem}{\begin{lemma}}
\newcommand{\elem}{\end{lemma}}
\newcommand{\bp}{\begin{problem}}
\newcommand{\ep}{\end{problem}}
\newcommand{\bn}{\begin{proposition}}
\newcommand{\en}{\end{proposition}}
\newcommand{\bd}{\begin{definition}}
\newcommand{\ed}{\end{definition}}
\newcommand{\bq}{\begin{question}}
\newcommand{\eq}{\end{question}}
\newcommand{\bprf}{\begin{proof}}
\newcommand{\eprf}{\end{proof}}
\newcommand{\br}{\begin{remark}}
\newcommand{\er}{\end{remark}}
\newcommand{\bs}{\begin{solution}}
\newcommand{\es}{\end{solution}}
\newcommand{\bnt}{\begin{noterm}}
\newcommand{\ent}{\end{noterm}}
\newcommand{\bnot}{\begin{notation}}
\newcommand{\enot}{\end{notation}}
\newcommand{\bterm}{\begin{terminology}}
\newcommand{\eterm}{\end{terminology}}

\newcommand{\id}{\mathrm{id}}

\newcommand{\Tr}{{\rm Tr} }

\def\R{{{\mathbb R}}}

\DeclareMathAlphabet{\mathpzc}{OT1}{pzc}{m}{it} 
 \DeclareFontFamily{OT1}{pzc}{}
 \DeclareFontShape{OT1}{pzc}{m}{it}{ <-> s*[1.2] pzcmi7t }{}
 \DeclareMathAlphabet{\mathpzc}{OT1}{pzc}{m}{it}

\def\M{\bold{M}}

\def\H{\mathcal{H}}

\def\P{\bold{P}}

\def\R{\mathbb{R}}

\def\D{\mathfrak{D}}

\newcommand{\stoch}{\;\xy0;/r.25pc/:(-3,0)*{}="1";(3,0)*{}="2";{\ar@{~>}"1";"2"|(1.06){\hole}};\endxy\!}

\title{The Born rule as a natural transformation of functors}
\author{Boyu Yang}
\affiliation{School of Mathematics and Statistics, Hainan University, Haikou, Hainan, 570228, China}
\author{James Fullwood}
\affiliation{School of Mathematics and Statistics, Hainan University, Haikou, Hainan, 570228, China}
\email{fullwood@hainanu.edu.cn}

\newcommand{\Addresses}{{
}}

\begin{document}
\emergencystretch 2em

\maketitle


\vspace{-7mm}
\tableofcontents

\begin{abstract}
In this work, we show that the quantum mechanical notions of density operator, positive operator-valued measure (POVM), and the Born rule, are all simultaneously encoded in the categorical notion of a \emph{natural transformation of functors}. In particular, we show that given a fixed quantum system $A$, there exists an explicit bijection from the set of density operators on the associated Hilbert space $\H_A$ to the set of natural transformations between the canonical measurement and probability functors associated with the system $A$, which formalize the way in which quantum effects (i.e., POVM elements) and their associated probabilities are additive with respect to a coarse-graining of measurements. 
\end{abstract}

	\maketitle

\section{Introduction}

Category theory makes mathematically precise the concepts of being and becoming, which perhaps are the most fundamental concepts in all of physics. In particular, a category consists of a collection of \emph{objects}---which may be thought of as all possible states of being associated with a particular type of structural entity---and a collection of \emph{morphisms}---which may be thought of as a class of admissible transformations between the various states of being. While category theory was first introduced in the context of pure mathematics~\cite{Eilenberg_45,Ma98}, it has since been utilized across a wide number of disciplines, including computer science, linguistics, neuroscience, and philosophy. In the context of quantum physics, category theory is central to the construction of topological quantum field theories~\cite{Atiyah88}, homological mirror symmetry in string theory~\cite{Kont_94}, topos-theoretic approaches to quantum foundations~\cite{D_ring_2010,Heunen_2012,Schreiber_2025}, a systematic study of Feynman diagrams~\cite{Baez_2001}, conformal field theory~\cite{Segal1988,schotten97}, higher gauge theory and quantum gravity~\cite{Baez_2010,Radenkovic_2020,Nikolic_2023}, the mysterious connection between path integrals and multiple zeta-values~\cite{Marcolli_2010}, quantum Bayesian inference and retrodiction~\cite{PaBayes,PaBu22,FuPa22a,Staton_2023,Fritz_2025}, and a diagrammatic formulation of quantum information-theoretic protocols~\cite{AbCo04,Fullwood_2025a}.

Despite the amount of interest in category theory in the context of quantum physics, it seems to have been overlooked that the basic constituents of quantum theory itself---namely, the notions of quantum state, measurement, and the Born rule---are all simultaneously encoded by the categorical notion of a \emph{natural transformation of functors}, as we show in this work. A functor is a mapping between categories that respects all of the categorical structure, i.e., it sends objects to objects and morphisms to morphisms. As such, a functor formalizes the notion of analogy, as it translates statements in one category to analogous statements in another category. A natural transformation then formalizes the notion of a mapping between functors, and is arguably the most fundamental notion of category theory. 

Here we consider the \emph{measurement} and \emph{probability functors}---which capture the way in which POVM elements and probabilities are additive with respect to a coarse-graining of measurements---and prove that a quantum state uniquely determines a natural transformation between the measurement and probability functors via the Born rule. Moreover, we prove that every such natural transformation between the measurement and probability functors is induced by the Born rule associated with a unique quantum state. 

In more precise mathematical terms, we prove Theorem~\ref{MTX1}, which states that given a fixed quantum system $A$, there exists a bijection from the set of density operators on the associated Hilbert space $\H_A$ to the set of natural transformations between the measurement and probability functors. To prove surjectivity, we make use of the Busch-Gleason Theorem~\cite{Gleason_57,Busch_2003}, which states that every generalized probability measure on the \emph{space of effects} (i.e., POVM elements on $A$) is induced by the Born rule associated with a unique density operator $\rho$ on $\H_A$. Thus, there is a precise sense in which Theorem~\ref{MTX1} may be viewed as a categorical lift of the Busch-Gleason Theorem. 

As a consequence of Theorem~\ref{MTX1}, it follows that once a Hilbert space $\H_A$ is associated with a quantum system $A$, the traditional separate postulates for quantum states, measurement, and the Born rule become redundant. In particular, these concepts are all subsumed within the single notion of a natural transformation between the measurement and probability functors associated with $\H_A$. Furthermore, such natural transformations inherently describe the dynamics of quantum effects and their associated probabilities in relation to the coarse-graining of measurements—a central component in the emerging theories of process matrices/tensors~\cite{OCB12,Shrapnel_2017,Pollock_2018} and quantum combs~\cite{chiribella_2008}. As such, our results provide yet another example of category theory's power in unifying core physical concepts.

In what follows, we recall the basic notions needed to define a natural transformation of functors in Section~\ref{S2}, and we give some illustrative examples. In particular, we go over the basic construction of the category $\mathfrak{Meas}$ consisting of measurable functions between measurable spaces, which plays a fundamental role in this work. In Section~\ref{S3} we introduce the measurement and probability functors---which are functors from the category $\mathfrak{Meas}$ to the category $\mathfrak{Set}$ of functions between sets---and show how such functors capture the fact that quantum effects and their associated probabilities are additive with respect to a coarse-graining of measurements. In Section~\ref{S4}, we prove that a natural transformation between the measurement and probability functors uniquely determines a generalized probability measure on the space of effects of a quantum system $A$, thus establishing a direct connection with the Busch-Gleason Theorem. In Section~\ref{S5}, we show how a density operator uniquely determines a natural transformation from the measurement functor to the probability functor via the Born rule, thus setting the stage for the proof of our main result Theorem~\ref{MTX1}.

\section{Categories, functors, and natural transformations} \label{S2}

In this section we provide the basic definitions of \emph{category}, \emph{functor}, and \emph{natural transformation}, while also providing some simple examples to help illustrate the concepts.
 
\bd \label{DX1}
A \define{category} $\mathscr{C}$ consists of the following data:
\begin{itemize}
\item
a class of objects, denoted by $\text{ob}(\mathscr{C})$.
\item
a class of morphisms, denoted by $\text{mor}(\mathscr{C})$.
\item
a function $\bold{dom}:\text{mor}(\mathscr{C})\to \text{ob}(\mathscr{C})$ given by $\bold{dom}(X\to Y)=X$ for every morphism $X\to Y$ in $\mathscr{C}$.
\item
a function $\bold{cod}:\text{mor}(\mathscr{C})\to \text{ob}(\mathscr{C})$ given by $\bold{cod}(X\to Y)=Y$ for every morphism $X\to Y$ in $\mathscr{C}$.
\item for every triple of objects 
$(X,Y,Z)$, a binary operation 
$\text{Hom}(X,Y) \times \text{Hom}(Y,Z)\to \text{Hom}(X,Z)$, which is referred to as \define{composition of morphisms}. Here, $\text{Hom}(X,Y)$ denotes the subclass of morphisms $f:X\to Y$ in $\text{mor}(\mathscr{C})$ such that $\text{dom}(f)=X$ and $\text{cod}(f)=Y$, and similarly for $\text{Hom}(Y,Z)$ and $\text{Hom}(X,Z)$. Given $(f,g)\in \text{Hom}(X,Y) \times \text{Hom}(Y,Z)$, the composition of $f$ and $g$ will be denoted by $g\circ f$. 
\end{itemize}
such that the following axioms hold:
\begin{itemize}
    \item (Associativity) If $f:X\to Y,g:Y\to Z$ and $h:Z\to W $ then $h\circ(g\circ f)=(h\circ g)\circ f$
    \item (Existence of Identity Morphisms) For every object $X\in \text{ob}(\mathscr{C})$, there exists a morphism $\id_X:X\to X$ such that for every morphism $f:X\to Y$, $f\circ\id_X=\id_Y\circ f=f$.
\end{itemize}
\ed

The prototypical example of a category is the category $\mathfrak{Set}$, whose objects are sets and whose morphisms consist of functions between sets. Many of the most ubiquitous categories are a refinement of $\mathfrak{Set}$ in the sense that the objects are sets with some extra structure, with the morphisms being functions which respect this structure, as in the following example. 

\bx
The category of partially ordered sets, denoted by $\mathfrak{Pos}$, is a fundamental example in category theory. It is defined as follows:
\begin{itemize}
    \item The objects of the category $\mathfrak{Pos}$ are all \textbf{partially ordered sets}, or \textbf{posets}. A poset is a set $P$ together with a binary relation $\le$ that is reflexive, antisymmetric, and transitive. Such a poset will be denoted by the pair $(P, \le)$.
    \item The morphisms in the category $\mathfrak{Pos}$ are the order-preserving maps. Given posets $(P,\leq_{P})$ and $(Q,\leq_Q)$, an \define{order-preserving map} consists of a function $f: P \to Q$ such that
\[
x \le_P y, \implies f(x) \le_Q f(y)\, .
\]
\end{itemize}
Moreover, a poset $(P,\le)$ itself may be thought of as a category, with the objects being the elements of $P$, and a single morphism from $x$ to $y$ if $x\leq y$, and no morphism from $x$ to $y$ otherwise.
\ex

In the next example we introduce the category of measurable functions between measurable spaces, which plays a fundamental role in this work.

\bx[The Category of Measurable Spaces]
A \define{measurable space} consists of a pair $(X,\Sigma_X)$, where $X$ is a set and $\Sigma_X$ is a $\sigma$-algebra of subsets of $X$, whose elements will be referred to as \define{events}. As the space of events $\Sigma_X$ is a $\sigma$-algebra, it is required to satisfy the following properties:
\begin{itemize}
\item
$X\in \Sigma_X$.
\item
$E\in \Sigma_X\implies X\setminus E\in \Sigma_X$.
\item If $\{E_n\}_{n=1}^{\infty}$ is a countable collection of sets in $\Sigma_{X}$, then their union $\bigcup_{n=1}^{\infty} E_n \in \Sigma_{X}$.
\end{itemize}

Given measurable spaces $(X,\Sigma_X)$ and $(Y,\Sigma_Y)$, a function $f:X\to Y$ is said to be \define{measurable with respect to $\Sigma_X$ and $\Sigma_Y$} if and only if $f^{-1}(F)\in \Sigma_X$ for all $F\in \Sigma_Y$. The category $\mathfrak{Meas}$ is then the category whose objects are measurable spaces, and a morphism $(X,\Sigma_X)\to (Y,\Sigma_Y)$ consists of a function $f:X\to Y$ which is measurable with respect to $\Sigma_X$ and $\Sigma_Y$. A measurable function $f:X\to Y$ with respect to $\Sigma_X$ and $\Sigma_Y$ will simply be referred to as \define{measurable} when the spaces of events $\Sigma_X$ and $\Sigma_Y$ are understood from the context. It is straightforward to show that $\mathfrak{Meas}$ indeed satisfies all the requirements of Definition~\ref{DX1}, and hence is a category.
\ex

\bd \label{DX2}
Let $\mathscr{C}$ and $\mathscr{D}$ be categories. A \define{(covariant) functor}  is a mapping $\bold{F}:\mathscr{C}\to \mathscr{D}$ that
\begin{itemize}
    \item
    associates each object $X\in \text{ob}(\mathscr{C})$ with an object $\bold{F}(X)\in \text{ob}(\mathscr{D})$,
    \item 
    associates each morphism $f:X\to Y$ in $\mathscr{C}$ to a morphism $\bold{F}(f):\bold{F}(X)\to \bold{F}(Y)$ in $\mathscr{D}$ such that the following two conditions hold:
    \begin{itemize}
    \item for every object $X\in \text{ob}(\mathscr{C})$, $\bold{F}(\id_X)=\id_{\bold{F}(X)}$.
    \item  for all morphisms $f:X\to Y$ and $g:Y\to Z$ in $\mathscr{C}$, $\bold{F}(g\circ f)=\bold{F}(g)\circ \bold{F}(f)$.
\end{itemize}
\end{itemize}
A \define{natural transformation} from a functor $\bold{F}:\mathscr{C}\to \mathscr{D}$ to a functor $\bold{G}:\mathscr{C}\to \mathscr{D}$ is a mapping $\mathcal{N}:\bold{F}\to \bold{G}$ that associates every $X \in \text{ob}(\mathscr{C})$ with a morphism $\mathcal{N}_X:\bold{F}(X)\to \bold{G}(X)$ in $\mathscr{D}$, such that for every morphism $f:X\to Y$ in $\mathscr{C}$ we have $\bold{G}(f)\circ \mathcal{N}_X=\mathcal{N}_Y\circ \bold{F}(f)$, i.e., the following diagram commutes.
\begin{equation*}
\xy0;/r.25pc/:
(-17.5,12.5)*+{\bold{F}(X)}="1";
(-17.5,-12.5)*+{\bold{F}(Y)}="2";
(17.5,12.5)*+{\bold{G}(X)}="0";
(17.5,-12.5)*+{\bold{G}(Y)}="3";
{\ar"1";"2"_{\bold{F}(f)}};
{\ar"1";"0"^{\mathcal{N}_X}};
{\ar"2";"3"_{\mathcal{N}_Y}};
{\ar"0";"3"^{\bold{G}(f)}};
\endxy
\end{equation*}
\ed

\bx[Group Action]
Let $G$ be a group. The category $\bold{B}G$ has a single object $\bullet$. The morphisms of $\bold{B}G$ are precisely the group elements of $G$, with composition given by group multiplication, i.e.,
for $g,h\in G$ we set $g\circ h=gh$. It then follows that a functor $\bold{F}:\bold{B}G\to \mathfrak{Set}$ consists of a set $X=\bold{F}(\bullet)$ together with a group of automorphisms of $X$, and thus corresponds to a \emph{group action} of $G$ on the set $X$. 
  
Now suppose we have two functors $\bold{F_1},\bold{F_2}:\mathbf{B}G\to\mathfrak{Set}$, so that $X=\bold{F_1}(\bullet)$ and $Y=\bold{F_2}(\bullet)$ are sets equipped with a group action of $G$, and suppose $\eta:\bold{F}_1\to \bold{F}_2$ is a natural transformation. It then follows that for all $g\in G$ the following diagram commutes:
\be \label{CMDXM57}
\xy0;/r.25pc/:
(-17.5,12.5)*+{X}="1";
(-17.5,-12.5)*+{X}="2";
(17.5,12.5)*+{Y}="0";
(17.5,-12.5)*+{Y}="3";
{\ar"1";"2"_{\bold{F_1}(g)}};
{\ar"1";"0"^{\eta}};
{\ar"2";"3"_{\eta}};
{\ar"0";"3"^{\bold{F_2}(g)}};
\endxy
\ee
Setting $\bold{F_1}(g)(x)=g\cdot x$ and $\bold{F_2}(g)(y)=g\cdot y$ for all $g\in G$, $x\in X$ and $y\in Y$, commutativity of diagram \eqref{CMDXM57} is equivalent to the condition
\[
\eta(g\cdot x)=g\cdot \eta(x) \qquad \forall g\in G,\;x\in X\, ,
\]
which is precisely the definition of $G$-equivariance of $\eta$. As such, natural transformations $\bold{F_1}\to \bold{F_2}$ are precisely the $G$-equivariant functions.
\ex

\section{The measurement and probability functors} \label{S3}

In this section, we introduce the \emph{measurement and probability functors}, which are functors from the category $\mathfrak{Meas}$ consisting of measurable functions between measurable spaces, to the category $\mathfrak{Set}$ consisting of functions between sets. To define the measurement and probability functors, we first need to set some notation and terminology. Throughout this work, we let $A$ denote a quantum system with a (separable) Hilbert space $\H_A$, which may be infinite-dimensional. The vector space of bounded self-adjoint operators on $\H_A$ will be denoted by $\bold{Obs}(A)$. The identity operator on $\H_A$ will be denoted by $\mathds{1}$, and the set of density operators---i.e., the set of positive, trace-class operators on $\H_A$ of unit trace---will be denoted by $\D(A)$. We note that for finite-dimensional systems, a density operator is simply a Hermitian operator on $\H_A$ whose finite set of eigenvalues are all non-negative and sum to 1. Given a measurable space $(X,\Sigma_X)$, a function $\mu:\Sigma_X\to \bold{Obs}(A)$ is said to be a \define{positive operator-valued measure (POVM)} if and only if $\mu$ satisfies the following properties:
\begin{itemize}
\item
$\mu(E)\geq 0$ for all $E\in \Sigma_X$.
\item
$\mu(X)=\mathds{1}$, where $\mathds{1}$ denotes the identity operator on $\H_A$.
\item
$\mu\left(\bigsqcup_{n=1}^{\infty}E_n\right)=\sum_{n=1}^{\infty}\mu(E_n)$ for every sequence of disjoint events $\{E_n\}$, where convergence is with respect to the ultra-weak operator topology on $\bold{Obs}(A)$.
\end{itemize}
A function $\mu:\Sigma_X\to \R$ is said to be a \define{probability measure} if and only if $\mu$ satisfies the same properties as above, but with the identity operator $\mathds{1}$ replaced by $1\in \R$.

\bd
The \define{measurement functor} is the mapping $\M:\mathfrak{Meas}\to \mathfrak{Set}$ defined by the following assignment.
\begin{itemize}
\item
\underline{On objects}: Given a measure space $(X,\Sigma_X)$, we let 
\[
\M(X,\Sigma_X)=\left.\Big\{\mu:\Sigma_X\to \bold{Obs}(A)\,\,\right|\,\,\text{$\mu$ is a POVM}\Big\}\, .
\]
\item
\underline{On morphisms}: Given a measurable function $f:X\to Y$, we let 
\[
\M(X\overset{f}\to Y):\M(X,\Sigma_X)\longrightarrow \M(Y,\Sigma_Y)
\]
be the function $\mu\mapsto f_*\mu$, where $(f_*\mu)(F)=\mu(f^{-1}(F))$ for all $F\in \Sigma_Y$.
\end{itemize}
\ed

\bd
The \define{probability functor} is the mapping $\P:\mathfrak{Meas}\to \mathfrak{Set}$ defined by the following assignment.
\begin{itemize}
\item
\underline{On objects}: Given a measurable space $(X,\Sigma_X)$, we let 
\[
\P(X,\Sigma_X)=\{\mu:\Sigma_X\to \R\,\,|\,\,\text{$\mu$ is a probability measure on $X$}\}\, .
\]
\item
\underline{On morphisms}: Given a measurable function $f:X\to Y$, we let 
\[
\P(X\overset{f}\to Y):\P(X,\Sigma_X)\longrightarrow \P(Y,\Sigma_Y)
\]
be the function $\mu\mapsto f_*\mu$, where $(f_*\mu)(F)=\mu(f^{-1}(F))$ for all $F\in \Sigma_Y$.
\end{itemize}
\ed

While it is well known to mathematicians that the measurement and probability functors are indeed functors, we nevertheless give a proof of this fact for the more general reader. 

\bn \label{FNCTXR67}
Tha mappings $\M:\mathfrak{Meas}\to \mathfrak{Set}$ and $\P:\mathfrak{Meas}\to \mathfrak{Set}$ are both functors.
\en

\bprf
We verify the functoriality conditions for $\M$. The verification for $\P$ then follows \emph{mutatis mutandis}. 

\vspace{0.5em}
\noindent
\textbf{(1) Identity preservation.}

Let $(X,\Sigma_X)$ be a measure space and let $\id_X : X \to X$ be the identity morphism. Then for all $\mu \in \M(X,\Sigma_X)$, and for all $E\in \Sigma_X$, we have
\[
\Big[\M(\id_X)(\mu)\Big](E)=({\id_X}_*\mu)(E)=\mu(\id^{-1}(E))=\mu(E)\, .
\]
It then follows that $\M(\id_X)(\mu)=\mu$ for all $\mu\in \M(X)$, thus $\M(\id_X) = \id_{\M(X)}$, as desired.

\vspace{0.5em}
\noindent
\textbf{(2) Composition preservation.}

Let $f : X \to Y$ and $g : Y \to Z$ be measurable functions. Then for all $\mu \in \M(X,\Sigma_X)$, and for all $G\in \Sigma_Z$, we have

\begin{align*}
\Big[\M(g \circ f)(\mu)\Big](G)&=\Big[(g\circ f)_*(\mu)\Big](G)=\mu\left((g\circ f)^{-1}(G)\right)=\mu\left(f^{-1}(g^{-1}(G))\right) \\
&=(f_*\mu)(g^{-1}(G))=\Big[g_*(f_*\mu)\Big](G)=\Big[\left(\M(g) \circ \M(f)\right)(\mu)\Big](G) \, ,\\
\end{align*}
thus $\M(g \circ f)=\M(g) \circ \M(f)$, as desired.

Since $\M$ preserves identity morphisms and preserves compositions, $\M$ is a functor, thus concluding the proof.
\eprf

The fact that $\M$ and $\P$ are functorial is a reflection of the fact that quantum effects (i.e., POVM elements) and their associated probabilities are additive with respect to a coarse-graining of measurements, and that this additivity is iterated through composition of measurable functions. In particular, suppose $(X,\Sigma_X)$ is a measurable space, let $E_1,E_2,\ldots$ be disjoint events in $\Sigma_X$, and suppose $M_i=\mu(E_i)$ for some POVM $\mu$. Now if $(Y,\Sigma_Y)$ is a measurable space and $f:X\to Y$ is a measurable function such that  
\[
f^{-1}(F)=\bigsqcup_i E_i
\]
for some $F\in \Sigma_Y$, then the event $F$ may be viewed as a coarse-graining of the events $E_i$. Moreover, it follows that the operator $N=(f_*\mu) (F)$ is such that 
\[
N=\sum_i E_i\, ,
\]
thus $N$ is a coarse-graining of the of the quantum effects $E_i$. As such, the POVM $f_*\mu$ may be viewed as a coarse-graining of the of the measurement $\mu$. Moreover, if $p_i=\Tr[\rho E_i]$ is the probability of the measurement outcome $E_i$ associated with some initial state $\rho\in \D(A)$ which is to be measured, then the probability of the effect $N$, namely, $\Tr[\rho N]$, is such that 
\[
\Tr[\rho N]=\sum_i p_i\, ,
\]
showing that probabilities of measurement outcomes are also additive with respect to a coarse-graining of measurements.

\section{Generalized probability measures on the space of effects} \label{S4}

In this section, we show that a natural transformation $\mathcal{N}:\M\to \P$ between the measurement and probability functors induces a generalized probability measure on the \define{space of effects} $\mathfrak{E}(A)\subset \bold{Obs}(A)$, which is the set given by
\[
\mathfrak{E}(A)=\{M\in \bold{Obs}(A)\,\,|\,\,0\leq M\leq \mathds{1}\}\, .
\] 
A function $\xi:\mathfrak{E}(A)\to [0,1]$ is said to be a \define{generalized probability measure} if and only if $\xi(\mathds{1})=1$, and for every countable set $\Lambda$ we have the implication
\be \label{ADTVXTY79}
\sum_{\lambda\in \Lambda}M_{\lambda}\leq \mathds{1}\implies \xi\left(\sum_{\lambda\in \Lambda}M_{\lambda}\right)=\sum_{\lambda\in \Lambda}\xi(M_{\lambda})\, .
\ee

A fundamental result in quantum theory is the Busch-Gleason Theorem~\cite{Gleason_57,Busch_2003}, which states that every generalized probability measure $\xi:\mathfrak{E}(A)\to [0,1]$ is of the form $\xi(M)=\Tr[\rho M]$ for some unique density operator $\rho\in \D(A)$. We now prove a lemma which will be crucial for the proof of our main result (Theorem~\ref{MTX1}), as it will put us in a position to apply the Busch-Gleason Theorem in the context of natural transformations between the measurement and probability functors.

\blem \label{LMXS1}
Let $\mathcal{N}:\M\to \P$ be a natural transformation, and let $\xi:\mathfrak{E}(A)\to [0,1]$ be the function given by 
\be \label{XINFDX17}
\xi(M)=\mathcal{N}_X(\mu)(E)
\ee
where $\mu:\Sigma_X\to \bold{Obs}(A)$ is any POVM such that $\mu(E)=M$. Then $\xi$ is a generalized probability measure on $\mathfrak{E}(A)$.\\
\elem

\bprf

\textbf{(1) Well-definedness.}
We first show that $\xi$ is well-defined, i.e. for any measurable space $(X,\Sigma_X)$ and $(Y,\Sigma_Y)$, POVMs $\mu \in \M(X,\Sigma_X)$, $\mu' \in \M(Y,\Sigma_Y)$, and $E \in \Sigma_X$, $F \in \Sigma_Y $ satisfying $\mu(E)=\mu'(F)=M $, we have
\[
\mathcal{N}_X(\mu)(E)
=
\mathcal{N}_{Y}(\mu')(F)\, .
\]
For this, let $(Z, \Sigma_Z)$ be the measurable space with $Z = \{z_1, z_0\}$ and $\Sigma_Z$ is its power set, and let $f:X\to Z$ and $g:Y\to Z$ be the measurable functions given by
\[
f(x) = 
\begin{cases} 
z_1 & \text{if } x \in E \\ 
z_0 & \text{if } x \notin E 
\end{cases} 
\quad \text{and} \quad
g(y) = 
\begin{cases} 
z_1 & \text{if } y \in F \\ 
z_0 & \text{if } y \notin F \, .
\end{cases}
\]
Now, consider their pushforward POVMs, $\nu = \M(f)(\mu)$ and $\nu' = \M(g)(\mu')$, which are defined on $(Z, \Sigma_Z)$. For the measurable set $\{z_1\} \in \Sigma_Z$, we have
\[
\nu(\{z_1\})= \mu(f^{-1}(\{z_1\})) = \mu(E) = M \quad \text{,} \quad \nu'(\{z_1\}) = \mu'(g^{-1}(\{z_1\})) = \mu'(F) = M\, ,
\]
and similarly for $\{z_0\} \in \Sigma_Z$ we have
\[
\nu(\{z_0\}) = \mu(f^{-1}(\{z_0\})) = \mu(X/E) = \mathds{1}-M \,\, \text{and} \,\, \nu'(\{z_0\}) = \mu'(g^{-1}(\{z_0\})) = \mu'(Y/F) = \mathds{1}-M\, ,
\]
thus $\nu = \nu'$. By naturality of $\mathcal{N}$, we then have
\[
\mathcal{N}_Z(\nu)({z_1})
=
\P(f)(\mathcal{N}_X(\mu))({z_1})
=
\mathcal{N}_X(\mu)(E)
\]
and
\[
\mathcal{N}_Z(\nu')({z_1})
=
\P(g)(\mathcal{N}_Y(\mu'))({z_1})
=
\mathcal{N}_Y(\mu')(F)\, .
\]
Since $\nu = \nu'$, it follows that
\[
\mathcal{N}_Z(\nu)({z_1})
=
\mathcal{N}_Z(\nu')({z_1})\, ,
\]
thus
\[
\mathcal N_X(\mu)(E)=\mathcal{N}_Y(\mu')(F)\, ,
\]
as desired.

\textbf{(2) Positivity and boundedness.}
Since for every measurable space $(X,\Sigma_X)$ and for every POVM $\mu:\Sigma_X\to \bold{Obs}(A)$ we have that $\mathcal{N}_X(\mu) \in \P(X,\Sigma_X)$ is a probability measure, it follows that $0 \le \xi(M) \le 1$.

\textbf{(3) Normalization.} Let $(X, \Sigma_X)$ be the measurable space with $X=\{\star\}$ and $\Sigma_X = \{\varnothing, X\}$. Then $\M(X,\Sigma_X)$ consists of a single POVM, namely, the POVM $\mu$ given by $\mu(X)=\mathds{1}$. By the definition of the natural transformation $\mathcal{N}$, $\mathcal{N}_X (\mu)$ is the unique probability measure in $\P(X,\Sigma_X)$ which assigns $1$ to $X$. We then have
\[
\xi(\mathds{1})=\mathcal{N}_X(\mu)(X)=1\, .
\]

\textbf{(4) Countable additivity.} We now show that the countable additivity condition \eqref{ADTVXTY79} holds for $\xi$. So let $\Lambda$ be a countable set, let $\{M_\lambda\}_{\lambda\in\Lambda}\subset\mathcal E(A)$ be a collection of effects satisfying
$\sum_{\lambda\in\Lambda} M_\lambda \le \mathds{1}$, and let $M=\sum_{\lambda\in \Lambda}M_{\lambda}$. Now define the discrete measurable space $(Z,\Sigma_Z)$, where $Z=\{0\}\sqcup\Lambda$ and $\Sigma_Z=2^Z$, and let $\mu: \Sigma_Z \to \bold{Obs}(A)$ be the function given by
\[
\mu(\{z\})=
\begin{cases}
M_{\lambda} & \text{if $z=\lambda\in \Lambda$} \\
1-M & \text{if $z=0$}\, .
\end{cases}
\]
Since $Z$ is countable, the above assignment extends to all of $\Sigma_Z$ by setting
\[
\mu(B)=\sum_{b\in B}\mu(\{b\}) \qquad \forall B\in \Sigma_Z\, .
\]
Since $M \le \mathds{1}$ we have $\mathds{1}-M \ge 0$, hence $\mu(\{0\})$ is an effect and $\mu(Z)=\mathds{1}$, hence $\mu$ is a POVM on $Z$. Now let $(Y, \Sigma_Y)$ be a simple two-outcome space where $Y = \{y_1, y_0\}$ and $\Sigma_Y=2^Y$ is its power set, let $f: Z \to Y$ be the measurable function given by
\[ 
f(j) = 
\begin{cases} 
y_1 & \text{if } j \in \Lambda \\ 
y_0 & \text{if } j = 0 \, ,
\end{cases} 
\]
and let $\nu = \M(f)(\mu)\in \P(Y, \Sigma_Y)$. We then have
\[ 
\nu(\{y_1\}) = \mu(f^{-1}(\{y_1\})) = \mu(\Lambda) = M \, , 
\]
and
\[
\nu(\{y_0\})=\mu(\{0\})=\mathds{1}-M\, .
\]
Moreover, by the naturality of $\mathcal{N}$ we have 
$\mathcal{N}_Y(\nu) = \P(f)(\mathcal{N}_Z(\mu))$. Now since
\[ 
\mathcal{N}_Y(\nu)(\{y_1\}) \stackrel{\text{def}}{=} \xi(\nu(\{y_1\})) = \xi(M)=\xi\left(\sum_{\lambda \in \Lambda}M_\lambda\right) \, ,
\] 
and
\[ 
\P(f)(\mathcal{N}_Z(\mu))(\{y_1\}) = \mathcal{N}_Z(\mu)(f^{-1}(\{y_1\})) = \mathcal{N}_Z(\mu)(\Lambda)=\sum_{\lambda \in \Lambda} \mathcal{N}_Z(\mu)(\lambda)=\sum_{\lambda \in \Lambda}\xi(M_{\lambda}) \, ,
\]
it follows that
\[ 
\xi\left(\sum_{\lambda \in \Lambda}M_\lambda\right) = \sum_{\lambda \in \Lambda} \xi(M_\lambda) \, ,
\]
as desired. This completes the proof that $\xi$ is a generalized probability measure on $\mathfrak{E}(A)$.
\eprf

\section{Quantum states as natural transformations} \label{S5}

In this section we show how a quantum state uniquely determines a natural transformation between the measurement and probability functors via the Born rule. Conversely, we show that \emph{every} natural transformation between the measurement and probability functors is induced from a unique quantum state. 

\bd
Given a density operator $\rho\in \D(A)$, let $\rho_*:\M\to \P$ be the mapping given by $(X,\Sigma_X)\mapsto \rho_*^X$, where $\rho_*^X:\M(X,\Sigma_X)\to \P(X,\Sigma_X)$ is the function given by
$\mu\mapsto \mu_{\rho}$, where
\[
\mu_{\rho}(E)=\Tr[\mu(E)\rho] \qquad \forall E\in \Sigma_X\, .
\]
\ed

\bn
The mapping $\rho_*:\M\to \P$ is a natural transformation of functors for all $\rho\in \D(A)$.
\en
\bprf
Let $\rho\in \D(A)$. We will prove naturality of $\rho_*$ in two parts.

\vspace{0.3em}
\noindent
\textbf{(1) Well-definedness on objects.}

Let $(X,\Sigma_X)$ be a measure space and let $\mu\in \M(X,\Sigma_X)$. We now show that $\mu_{\rho}=\rho_*^X(\mu)\in \P(X,\Sigma_X)$, showing that the map $\rho_*^X$ is well-defined for all measure spaces $(X,\Sigma_X)$. Indeed, since $\rho \ge 0$ and $\mu(E) \ge 0$ for all $E\in \Sigma_X$, it follows that 
\[
\Tr[\mu(E) \rho] \ge 0\qquad \forall E\in \Sigma_X\, .
\]
Moreover,
\[
\mu_\rho(X)
=
\Tr[\mu(X) \rho]
=
\Tr\left[ \mathds{1} \rho \right]
=
\Tr[\rho]
= 1\, ,
\]
thus $\mu_\rho \in \P(X,\Sigma_X)$, as desired.

\vspace{0.5em}
\noindent
\textbf{(2) Naturality condition.}

Let $f : X \to Y$ be a measurable function, and let $\mu \in \M(X,\Sigma_X)$. For all $F \in \Sigma_Y$ we then have
\begin{align*}
\Big[\rho_*^Y(\M(f)(\mu))\Big](F)&=
(\M(f)(\mu))_\rho(F)=
\Tr\left[ \M(f)(\mu)(F) \rho \right]=
\Tr\left[ \mu(f^{-1}(F)) \rho \right] \\
&=
\mu_{\rho}(f^{-1}(F))=
\Big[\P(f)(\mu_\rho)\Big](F)=
\Big[\P(f)(\rho_*^X(\mu))\Big](F)\, .
\end{align*}
Therefore, $\rho_*^Y \circ \M(f)=\P(f) \circ \rho_*^X$, thus $\rho_*$ is a natural transformation, as desired.
\eprf

Now let $\mathfrak{Nat}(\M,\P)$ denote the set of all natural transformations from $\M$ to $\P$. We now show that the mapping $\rho\mapsto \rho_*$ induces a bijective correspondence between the set of density operators $\D(A)$ and the set $\mathfrak{Nat}(\M,\P)$.

\bt \label{MTX1}
The mapping $\Phi:\D(A)\to \mathfrak{Nat}(\M,\P)$ given by $\Phi(\rho)=\rho_*$ is a bijection.
\et

\bprf
\textbf{Injectivity.} \noindent Let $\rho,\sigma\in \D(A)$ be two distinct density operators, and let $\Delta\in \bold{Obs}(A)$ be the self-adjoint operator given by $\Delta = \rho - \sigma\neq 0$. Clearly $\Tr[\Delta]=0$ as density operators are of unit trace. By the spectral theorem for self-adjoint operators, $\Delta$ may be written in terms of the spectral integral
\[
\Delta = \int_{\sigma(\Delta)} \lambda \, dE_\Delta(\lambda)\, ,
\]
where $E_\Delta$ is the unique projection-valued measure which maps Borel subsets of the spectrum $\sigma(\Delta)$ to projection operators on $\mathcal{H}_A$. Since $\Delta \neq 0$ and $\Tr(\Delta) = 0$, $\Delta$ can be neither positive semi-definite nor negative semi-definite, thus $\sigma(\Delta)$ must contain both positive and negative values. It then follows that there exists $\epsilon > 0$ such that the spectral projection corresponding to the interval $(\epsilon, \infty)$ is a non-zero operator, which we will denote by $P$. Using the functional calculus, it follows that the operator product $\Delta P$ is given by
\[
\Delta P = \left( \int_{\sigma(\Delta)} \lambda \, dE_\Delta(\lambda) \right) P = \int_{(\epsilon, \infty)} \lambda \, dE_\Delta(\lambda)\, .
\]
In this integral, the variable $\lambda$ is strictly positive over the entire domain of integration ($\lambda > \epsilon > 0$). Moreover, since the projection $P \neq 0$, the resulting operator $\Delta P$ is a non-zero, positive trace-class operator, thus 
\[
\Tr[\Delta P] > 0\implies \Tr[P\rho] \neq \Tr[P\sigma]\, .
\]

Now let $X = \{ x, x' \}$, let $\Sigma_X=2^X$, and let $\mu \in \M(X,\Sigma_X)$ be the POVM given by
\[
\mu(\{x\}) = P, \quad \mu(\{x'\}) = \mathds{1} - P\, .
\]
We then have
\[
\rho_{*}^X(\mu)(\{x\})=\Tr[ \mu(\{x\})\rho]
= \Tr[P\rho]
\neq
\Tr[P\sigma]=\Tr[ \mu(\{x'\})\sigma]
= \sigma_{*}^X(\mu)(\{x\})\, ,
\]
thus $\rho_{*} \neq \sigma_{*}$. This establishes the injectivity of $\Phi$.\\
\textbf{Surjectivity.} 
Let $\mathcal{N}\in\mathfrak{Nat}(\M,\P)$ be a natural transformation. By Lemma~\ref{LMXS1}, the function $\xi:\mathfrak{E}(A)\to [0,1]$ given by \eqref{XINFDX17} is a generalized probability measure, thus by the Busch–Gleason Theorem there exists a unique $\rho \in \D(A)$ such that 
\[
\xi(M)=\Tr[\rho M] \qquad \forall M\in \mathfrak{E}(A)\, .
\]
Now let $(X,\Sigma_X)$ be any measurable space, $\mu \in \mathsf \M(X)$ any POVM,
and $E \in \Sigma_X$ any event. We then have
\[
N_X(\mu)(E) = \xi(\mu(E)) = \Tr\big[\rho \,\mu(E)\big]
= \rho_*^X(\mu)(E)\, ,
\]
which implies $\mathcal{N} = \rho_*=\Phi(\rho)$.
This establishes the surjectivity of $\Phi$, thus concluding the proof.
\eprf

\addcontentsline{toc}{section}{\numberline{}Bibliography}
\bibliographystyle{quantum}
\bibliography{references}

\Addresses

\end{document}